\documentclass[sigplan,10pt]{acmart}
\renewcommand\footnotetextcopyrightpermission[1]{}
\AtBeginDocument{%
  }
\usepackage{graphicx}
\usepackage{subcaption}
\usepackage{algorithm}
\usepackage{algorithmic}
\usepackage{booktabs}
\usepackage{pifont}
\usepackage[most]{tcolorbox}
\usepackage{placeins}

\newtcolorbox{takeawaybox}{
    colback=gray!5,
    colframe=black!35,
    boxrule=0.4pt,
    arc=1pt,
    left=4pt,
    right=4pt,
    top=3pt,
    bottom=3pt,
    fontupper=\small,
    before skip=4pt,
    after skip=4pt
}

\setcopyright{acmlicensed}
\copyrightyear{2018}
\acmYear{2018}
\acmDOI{XXXXXXX.XXXXXXX}
\begin{document}

\title{HorizonServe: Coordinating Request Scheduling with GPU Sharing for Omni-Model Serving}

\author{Yuning Zhang}
\email{yuning.zhang1@sydney.edu.au}
\affiliation{%
  \institution{School of Electrical and Computer Engineering, The University of Sydney}
  \streetaddress{J03 Building, Maze Cres}
  \city{Sydney}
  \state{NSW}
  \country{Australia}
  \postcode{2008}
}

\author{Dong Yuan}
\email{dong.yuan@sydney.edu.au}
\authornote{Corresponding author: Dong Yuan.}
\affiliation{%
  \institution{School of Electrical and Computer Engineering, The University of Sydney}
  \streetaddress{J03 Building, Maze Cres}
  \city{Sydney}
  \state{NSW}
  \country{Australia}
  \postcode{2008}
}

% \affiliation{%

% \affiliation{%

% \affiliation{%

% \affiliation{%

% \affiliation{%

% \affiliation{%

% \affiliation{%


\begin{abstract}
Omni models unify text, speech, image, and multimodal reasoning in a single serving backend, but this unified deployment exposes a new scheduling problem. Requests with different output modalities may share an initial multimodal backbone and then diverge into downstream generation stages, creating heterogeneous first-response metrics and service-level objective (SLO) targets on the same GPU. Existing large language model (LLM) and multimodal serving systems mainly optimize token progress or input-side processing, and they do not jointly control temporal sharing in the shared stage and spatial sharing among co-running stages. This paper presents HorizonServe, a single-GPU omni-model serving system that coordinates request admission and GPU allocation under heterogeneous SLOs. HorizonServe profiles per-class first-response latency, protects requests with limited slack, rotates shared-stage opportunities across execution paths, and throttles the shared-stage streaming multiprocessor (SM) allocation when downstream stages are active. Across three omni-model workloads and two GPU platforms, HorizonServe improves SLO attainment by up to 4.9$\times$ in arrival-rate sweeps and 7.0$\times$ under downstream-heavy traffic, and reduces per-class first-response latency by 38.4--63.7\%.
\end{abstract}





\pagestyle{plain}
\maketitle
\section{Introduction}
\label{sec:introduction}

Interactive model services are moving from text-only chat toward omni models that accept and produce text, speech, image, and video. GPT-4o was an early public example of real-time reasoning across audio, vision, and text~\cite{openai2024gpt4o}, and recent open omni or unified multimodal models extend similar capabilities to text, speech, image, video, and multimodal generation~\cite{xu2025qwen25omnitechnicalreport,deng2025bagel,coreteam2025mimoaudioaudiolanguagemodels}. This changes the serving target. A serving system no longer hosts only a homogeneous text generator, but instead may host a multi-stage model that receives heterogeneous requests and routes them through different output paths.

Omni models are also useful beyond large public assistants. In private, on-premises, and edge settings, speech, images, and video can be sensitive, bandwidth-heavy, and latency-sensitive, motivating inference close to where data is produced~\cite{ruan2024webllm,chu2023mobilevlm,wang2025lmmeter,microsoft-local-ai,nvidia-edge-ai}. Large services can amortize omni-model execution across GPU clusters or disaggregate stages across devices~\cite{vllm,298679,yin2026vllmomnifullydisaggregatedserving}, but site-level deployments often have a smaller resource envelope. Recent omni models already support local inference, low-memory execution, and edge deployment~\cite{xu2025qwen25omnitechnicalreport}, while workstation GPUs are positioned for local generative-AI inference~\cite{nvidia-rtx-ai-workstations,nvidia-rtx6000-ada,nvidia-rtxpro6000-workstation}. This paper studies single-GPU omni-model serving under heterogeneous service-level objectives (SLOs).
\begin{figure}[!t]
    \centering
    \begin{subfigure}[t]{\columnwidth}
        \centering
        \includegraphics[width=0.9\columnwidth]{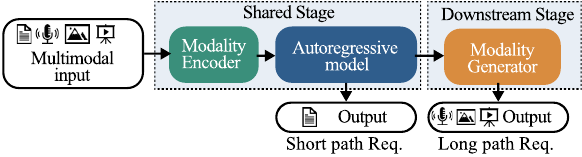}
        \caption{Omni-model execution architecture.}
        \label{fig:intro-omnimodel}
    \end{subfigure}
    \vspace{0.35em}
    \begin{subfigure}[t]{\columnwidth}
        \centering
        \includegraphics[width=0.9\columnwidth]{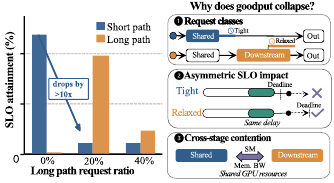}
        \caption{Goodput collapse as long-path traffic increases.}
        \label{fig:intro-sharing}
    \end{subfigure}
    \caption{Single-GPU omni-model serving couples heterogeneous request paths and GPU resource sharing.}
    \label{fig:intro}
\end{figure}

Single-GPU deployment may seem like a packaging constraint: fit all components on one GPU and reuse an existing serving system. The difficulty is that an omni-model service is not a homogeneous workload. Figure~\ref{fig:intro-omnimodel} sketches the execution architecture: requests first enter a shared multimodal stage, then either return text immediately or continue into output-specific generators, such as speech decoders, codec modules, image generators, or renderers~\cite{xu2025qwen25omnitechnicalreport,xu2025qwen3omnitechnicalreport,coreteam2025mimoaudioaudiolanguagemodels}. On a single GPU, this structure creates two coupled sharing problems: heterogeneous request paths share the initial stage over time, while co-running model stages share SMs and GPU memory-bandwidth resources in space.

This coupling breaks the assumptions behind text-only SLO scheduling. Different output modalities expose different first-response metrics, such as time-to-first-token (TTFT) for text and time-to-first-output (TTFO) for speech or image generation. Their natural response times are different by orders of magnitude, so the same shared-stage delay can be harmless for one class but fatal for another. Figure~\ref{fig:intro-sharing} shows this effect in a motivating workload: as long-path traffic increases, short-path requests lose SLO attainment first, which directly reduces SLO-satisfied goodput at the same offered load. The collapse is not simply a higher average load, and it reflects asymmetric SLO impact under shared-stage delay. Spatial contention reinforces the problem because downstream generators can slow the shared stage by consuming memory bandwidth and execution resources.

Prior systems cover adjacent axes of this problem but not their combination. LLM serving systems assume a common text-generation path~\cite{280922,vllm,298679,298687}; VLM/LMM serving mainly handles input-side multimodal processing before text generation~\cite{guo2025rserve,10.1145/3772052.3772254,bai2026epd}; and GPU sharing systems expose co-location or partitioning mechanisms without deciding which omni paths should enter the shared stage~\cite{nvidia-mps,nvidia-mig,nanoflow,bullet}. The missing interface is the one created by omni models: a scheduling decision at the shared stage can change both immediate SLO slack and later downstream GPU pressure. HorizonServe's contribution is the omni-specific feedback loop that connects output-path admission, deadline slack, downstream activation, and bounded stage-level GPU allocation on one GPU.

We present HorizonServe, a serving system for single-GPU omni-model workloads with heterogeneous first-response SLOs. HorizonServe coordinates two decisions at each scheduling opportunity: which execution paths should enter the shared stage, and how much GPU parallelism the shared stage should use when downstream stages are active. It realizes these decisions with three mechanisms. First, HorizonServe profiles per-class first-response latency and estimates remaining slack from offline stage profiles and online progress, so near-deadline requests can bypass ordinary path filtering. Second, it uses shared-stage path rotation with a long-path admission cap, preventing long-path bursts from monopolizing shared-stage slots and limiting future downstream activation. Third, it maps runtime memory-bandwidth pressure to a bounded shared-stage SM cap during co-running, giving downstream stages headroom while avoiding the harmful full-allocation regime. Together, these mechanisms turn uncontrolled temporal and spatial sharing into explicit SLO-oriented control.

This paper makes the following contributions:
\begin{itemize}
    \item We identify a new serving problem in single-GPU omni-model deployment. Heterogeneous omni requests share an initial stage in time and co-located downstream stages share the same GPU in space, causing asymmetric SLO violations and cross-stage interference.

    \item We design and implement HorizonServe, an inference serving system for omni-model workloads with heterogeneous first-response SLOs. HorizonServe jointly schedules shared-stage admission and stage-level GPU allocation using request slack and runtime bandwidth pressure.

    \item We evaluate HorizonServe across omni-model workloads and GPU platforms against representative serving baselines. HorizonServe improves SLO attainment by up to 4.9$\times$ in arrival-rate sweeps and 7.0$\times$ under downstream-heavy traffic, while reducing per-class first-response latency by 38.4--63.7\%.
\end{itemize}

\section{Background}
\label{sec:background}

\subsection{Model Components and Execution Patterns}
\label{sec:bg-model-components}
Modern omni-models combine several components for multimodal understanding and generation. We briefly review the building blocks that shape serving behavior: modality encoders, autoregressive transformers, visual generators, and audio-codec-based speech generators.

\textbf{Modality encoders.} Omni-models often convert images, video frames, or audio features into representations consumed by the multimodal backbone. Representative image encoders include CLIP and SigLIP~\cite{radford2021clip,zhai2023siglip}. Whisper is a representative audio encoder that maps waveform inputs into speech representations~\cite{radford2023robust}. Because modality encoders are usually lightweight, serving systems often co-execute them with the AR stage instead of treating them as independent serving stages.

\textbf{Autoregressive transformers.} Autoregressive (AR) transformers generate tokens sequentially~\cite{NIPS2017_3f5ee243,radford2018improving} and are widely used for language reasoning, multimodal understanding, and token generation in recent omni or unified multimodal models~\cite{xu2025qwen3omnitechnicalreport,xu2025qwen25omnitechnicalreport,deng2025bagel,coreteam2025mimoaudioaudiolanguagemodels,wu2024janusdecouplingvisualencoding,chen2025janusprounifiedmultimodalunderstanding}. AR inference consists of prefill, which processes the input context and builds the key-value (KV) cache, and decode, which repeatedly accesses model weights and the accumulated KV cache to generate new tokens.

\textbf{Visual generators.} Diffusion models~\cite{rombach2022high,NEURIPS2020_4c5bcfec} and diffusion transformers (DiTs)~\cite{peebles2023scalable} generate visual content through iterative denoising or refinement. They appear in recent unified omni-models~\cite{deng2025bagel,wu2024janusdecouplingvisualencoding} and create output-side generation loops distinct from AR decode.

\textbf{Audio-codec and waveform generators.} Speech-output omni-models often generate or consume discrete audio codes before producing waveform samples. Recent audio-capable omni-models~\cite{coreteam2025mimoaudioaudiolanguagemodels,meituanlongcatteam2025longcatflashomnitechnicalreport} use speech-token generators, codec decoders, residual vector quantization (RVQ), or code-to-waveform modules~\cite{zeghidour2021soundstream,defossez2022high,10842513,kong2020hifi}, adding generation or reconstruction loops beyond text decoding.

\subsection{Omni-Model Serving Patterns}
\label{sec:bg-omni-architectures}
Omni models are different from conventional multimodal models at the model boundary. Many vision-language and large multimodal serving targets accept non-text inputs but still expose a single text-generation path: images, audio, or video are encoded on the input side and then consumed by an autoregressive language model~\cite{guo2025rserve,10.1145/3772052.3772254,bai2026epd}. Recent omni and unified multimodal models extend the output side as well. A single model family may return text directly, synthesize speech through a Talker or codec decoder, generate images through visual experts, or combine several generation modules~\cite{xu2025qwen25omnitechnicalreport,deng2025bagel,coreteam2025mimoaudioaudiolanguagemodels}. Thus, an omni model is not merely a text LLM with richer inputs. From the serving perspective, it is a coupled multi-stage model with conditional output paths.

Table~\ref{tab:omni-model-taxonomy} summarizes representative models and the serving patterns we abstract from them. Their architectures are different: some separate a reasoning backbone from a speech generator, some attach image generation experts, and others fuse audio-language modeling with waveform reconstruction. The common feature is that requests first share an initial multimodal computation, but only some requests continue into output-specific generation.

\begin{table}[!t]
\centering
\scriptsize
\setlength{\tabcolsep}{2.5pt}
\caption{
Representative omni and unified multimodal models. T, S, and I denote text, speech, and image output. S gen, I gen, and multi-gen denote downstream speech, image, and multiple generation paths after the shared multimodal computation.
}
\label{tab:omni-model-taxonomy}
\begin{tabular*}{\columnwidth}{@{}l@{\extracolsep{\fill}}ccc@{}}
\toprule
\textbf{Model} & \textbf{Output} & \textbf{Org.} & \textbf{Serving pattern} \\
\midrule
Qwen2.5-Omni~\cite{xu2025qwen25omnitechnicalreport}
& T,S & Thinker-Talker & Shared + S gen \\

Qwen3-Omni~\cite{xu2025qwen3omnitechnicalreport}
& T,S & Thinker-Talker & Shared + S gen \\

LongCat-Flash-Omni~\cite{meituanlongcatteam2025longcatflashomnitechnicalreport}
& T,S & LLM + speech & Shared + S gen \\

MiniCPM-o~\cite{yao2024minicpm}
& T,S & Omni LLM & Shared + S gen \\

VITA-1.5~\cite{fu2026vita}
& T,S & Omni LLM & Shared + S gen \\

InteractiveOmni~\cite{tong2025interactiveomniunifiedomnimodalmodel}
& T,S & LLM + speech & Shared + S gen \\

Ming-Omni~\cite{ai2025mingomniunifiedmultimodalmodel}
& T,S,I & MoE / routed & Shared + multi-gen \\

Ming-Flash-Omni~\cite{ai2026mingflashomnisparseunifiedarchitecture}
& T,S,I & Sparse / routed & Shared + multi-gen \\

BAGEL~\cite{deng2025bagel}
& T,I & MoT experts & Shared + I gen \\

MiMo-Audio~\cite{coreteam2025mimoaudioaudiolanguagemodels}
& T,S & Fused audio LM & Shared + waveform \\
\bottomrule
\end{tabular*}
\end{table}

The table highlights a model-level serving shape behind these systems. A request first enters a shared backbone, expert, or fused thinker-talker that performs input understanding, cross-modal reasoning, token generation, or routing. We call this component the \textbf{shared stage}. The shared stage also includes input-side modality encoders unless the serving runtime explicitly separates them, because these encoders are part of the computation that corresponding multimodal requests must pass through before output generation begins.

After the shared stage, requests diverge. Text-output requests can often return their first visible response directly from the shared stage. Speech-output requests may enter a Talker, audio-code generator, codec decoder, or Code2Wav module before the first audio chunk is available. Image-output requests may enter a diffusion model, a visual generator, or a generation expert before a rendered image is produced. We call these output-side components the \textbf{downstream stage}. They are not always active: a short text request may never use them, while a speech or image request can activate one or more of them after consuming shared-stage work.

This abstraction is serving-oriented rather than architecture-specific. A Thinker--Talker model, a mixture-of-transformers model, and a fused audio language model can all be viewed as sharing an initial stage and optionally invoking output-side generation. We call requests that stop at the shared stage \textbf{short-path} requests, and requests that continue downstream \textbf{long-path} requests. This distinction is central to HorizonServe because existing text and input-side multimodal serving abstractions do not expose these output-path choices: short and long paths contend for the same shared-stage scheduling opportunities, but create different downstream activity and first-response SLOs.

\subsection{SLO-Oriented Serving and GPU Sharing}
\label{sec:bg-slo-serving}

Online model serving systems are commonly evaluated using latency targets or service-level objectives (SLOs). For text generation, a key first-response metric is time-to-first-token (TTFT). Omni-model serving adds first-response metrics for paths that terminate at different stages, such as time-to-first-output (TTFO) for the first audio chunk, speech segment, or rendered image. Since these outputs have different natural time scales, each request class may need its own SLO target.

Our optimization goal is SLO-satisfied goodput: useful throughput from requests whose first-response latency satisfies a class-specific SLO. Goodput is widely used in serving systems that optimize latency under load~\cite{298687,298679,chen2026towards}. We report SLO attainment, the fraction of requests that meet the corresponding target, because omni-model requests expose different first-response metrics and SLO slack.

These metrics distinguish omni-model serving from text-only serving. LLM serving usually optimizes a prefill--decode loop around TTFT, token latency, or online goodput~\cite{298687,298679,280922}. VLM and LMM serving systems add input-side encoders and multimodal preprocessing before text generation~\cite{guo2025rserve,10.1145/3772052.3772254,bai2026epd}. Omni-model serving must also attach SLOs to output-side generation paths, where the first visible response may be a token, audio chunk, or image.

GPU runtimes provide mechanisms for spatial sharing. CUDA streams expose concurrent work submission within a process, while NVIDIA Multi-Process Service (MPS) allows kernels from multiple processes to execute concurrently on one GPU~\cite{nvidia-cuda-programming-guide,nvidia-mps}. Multi-Instance GPU (MIG) partitions supported GPUs into isolated instances~\cite{nvidia-mig}. For finer-grained SM control, libsmctrl exposes hardware compute partitioning, and CUDA Green Contexts provide a driver API for provisioning SM resources to lightweight contexts~\cite{bakita2023hardware,nvidia-green-contexts}. These mechanisms motivate treating the shared-stage SM count as a runtime control knob.

\section{Motivation}
\label{sec:motivation}

% The following experiments are meant to expose the scheduling effects that appear in single-GPU omni-model serving. We use solo p95 first-response latency to derive each request class's SLO target, report SLO attainment, and keep the total arrival process fixed when changing the traffic mix.

\subsection{SLO Heterogeneity}
\label{sec:motivation-slo}

\begin{figure}[!t]
    \centering
    \includegraphics[width=0.8\columnwidth]{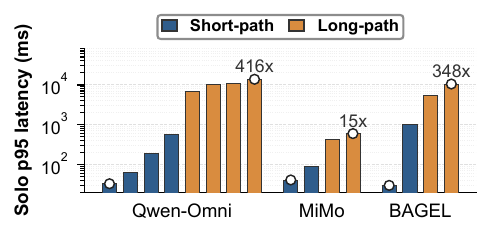}
    \caption{
    Solo p95 first-response latency across request types on one GPU. Short-path and long-path requests occupy distinct latency regimes across Qwen-Omni, MiMo-Audio, and BAGEL. 
    }
    \label{fig:m1-slo-spread}
\end{figure}

Omni-model serving multiplexes requests that live in fundamentally different latency regimes. A text-output request may complete its first response after only the shared stage, while a speech or image generation request continues through downstream generators before producing user-visible output. Even when each request type runs alone on one GPU, these paths already exhibit orders-of-magnitude gaps in p95 first-response latency, as shown in Figure~\ref{fig:m1-slo-spread}. This gap is not merely a performance detail. In online serving, SLO targets are typically derived from each request class's solo latency, so the latency spread becomes an SLO spread: some requests have sub-second deadlines, while others naturally tolerate second-level first-response latency.

This heterogeneity changes the meaning of sharing. If all request classes are scheduled through the same shared stage, a delay that is negligible for a long-path request can consume a large fraction of the slack of a short-path request. Thus, a single-GPU omni server must not only maximize aggregate throughput; it must preserve latency slack across request classes whose natural response times are separated by orders of magnitude.

\begin{takeawaybox}
\textbf{Takeaway 1.} Omni serving starts from heterogeneous latency regimes, so shared-stage scheduling must preserve SLO slack rather than treating all request classes as interchangeable work.
\end{takeawaybox}
\subsection{Goodput Collapse}
\label{sec:motivation-collapse}

We next show that heterogeneous serving paths can cause severe goodput loss even when long-path requests account for only a small fraction of the workload. We vary the fraction of long-path traffic from 0\% to 60\%. For Qwen-Omni, long-path requests generate audio; for BAGEL, they generate images. Short-path requests in both workloads produce text and terminate after the shared stage. Since the offered load is fixed within this sweep, a drop in SLO attainment corresponds to a drop in SLO-satisfied goodput.

\begin{figure}[!t]
    \centering
    \includegraphics[width=0.8\columnwidth]{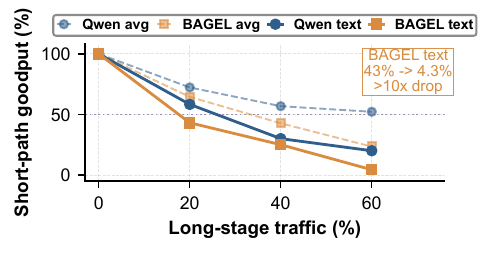}
\caption{
Long-path traffic collapses short-path goodput under single-GPU co-location.
We plot SLO attainment, which is the normalized SLO-satisfied goodput at fixed offered load.
The tightest-SLO class degrades first as long-path traffic increases.
}
    \label{fig:motivation-collapse}
\end{figure}

% In Qwen-Omni, introducing only 20\% long-path traffic reduces the attainment of the text request from 100\% to 58.3\%. In BAGEL, the text request is even more sensitive: its attainment drops from 43\% at 20\% long-path traffic to 4.3\% at 60\%, a more than 10$\times$ reduction.
Figure~\ref{fig:motivation-collapse} shows that short-path SLO attainment, and hence normalized goodput, drops sharply as long-path traffic increases. The solid lines show the tightest-SLO text request, while the dashed lines show aggregate attainment across all short-path request types. The aggregate short-path curves remain higher than the text-only curves because other short-path request types have looser SLO targets, but they also decline as long-path traffic increases.

This result shows that long-path traffic does not merely add proportional load. Long-path requests share the same initial stage with short-path requests before continuing to downstream generators, so they consume shared-stage scheduling opportunities before producing their own downstream work. The resulting delay is experienced by multiple short-path classes, but its SLO impact is asymmetric: a small increase in shared-stage waiting time can consume most of the slack of a tight-SLO request, while looser classes may still complete before their deadlines. This motivates a shared-stage scheduler that regulates which paths enter the shared stage while still allowing near-deadline requests to bypass normal ordering.
\begin{takeawaybox}
\textbf{Takeaway 2.} Long-path traffic can sharply reduce short-path goodput, and the tightest-SLO request class degrades first because shared-stage delay consumes its limited slack.
\end{takeawaybox}

\subsection{Cross-Stage Bandwidth Contention}
\label{sec:motivation-spatial}
\begin{figure}[!t]
    \centering
    \begin{subfigure}[t]{\columnwidth}
        \centering
        \includegraphics[width=0.85\linewidth]{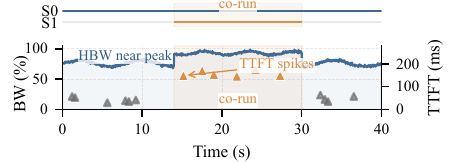}
        \caption{Cross-stage co-running over time.}
        \label{fig:m4a-timeline}
    \end{subfigure}

    \begin{subfigure}[t]{\columnwidth}
        \centering
        \includegraphics[width=0.85\linewidth]{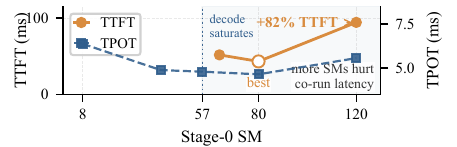}
        \caption{SM allocation, shared-stage scaling, and short-request TTFT.}
        \label{fig:m4b-sm}
    \end{subfigure}

    \caption{
    Cross-stage bandwidth contention in single-GPU omni-model serving.
    }
    \label{fig:motivation-spatial}
\end{figure}

Temporal interference alone does not fully explain the degradation in mixed omni-model workloads. In the single-GPU co-location setting, long-path requests that continue beyond the shared stage also activate downstream generators on the same GPU. As a result, the shared stage and downstream stage do not run in isolation: they execute concurrently and compete for the same physical resources. Figure~\ref{fig:motivation-spatial} is a fixed-SM profiling experiment rather than a controller trace: we use CUDA Green Contexts to sweep the shared-stage SM allocation and expose when full shared-stage allocation becomes unsafe under co-running.

Figure~\ref{fig:motivation-spatial}(a) visualizes this process over time. Shared-stage decode remains active throughout the execution, while the downstream stage becomes active only during the co-running interval. Before downstream execution starts, the measured GPU memory-bandwidth utilization stays around 76\% of peak and short text-output requests show stable TTFTs. Once the downstream stage becomes active, memory-bandwidth utilization rises to about 93\% of peak, and the TTFT of short text-output requests frequently spikes into the 90--200\,ms range. This temporal alignment indicates that downstream execution increases memory pressure on the shared GPU, which in turn degrades the first-response latency of short requests.

Figure~\ref{fig:motivation-spatial}(b) further diagnoses the source of this interference. The dashed curve shows the isolated scaling behavior of shared-stage decode under fixed SM allocations. Decode TPOT is already close to saturation in the 64--80 SM range: it reaches 4.80\,ms at 64 SM and 4.65\,ms at 80 SM, but worsens to 5.57\,ms at 120 SM. This indicates that full shared-stage allocation provides no TPOT benefit once decode approaches its useful saturation range. In contrast, the solid curve shows the TTFT of short text-output requests when the shared stage and downstream stage co-run. Allocating 120 SM to the shared stage increases TTFT from 43.3\,ms at 80 SM to 94.2\,ms, a 118\% degradation. Thus, giving the shared stage a full allocation under co-running can amplify cross-stage bandwidth contention and harm first-response latency.

This result clarifies an important design point. The dominant cross-stage bottleneck is not simply SM shortage, but GPU memory-bandwidth contention. Full shared-stage allocation is unsafe under co-running because it can increase memory pressure without improving decode TPOT. This observation motivates bounding the shared-stage cap near its useful saturation range and leaving the remaining SMs as downstream headroom, rather than allowing the shared stage to expand to the full GPU allocation by default.

\begin{takeawaybox}
\textbf{Takeaway 3.} In single-GPU omni-model serving, full shared-stage allocation is unsafe under co-running. The shared stage should be bounded near its useful saturation range to reduce cross-stage bandwidth contention and protect short-request latency.
\end{takeawaybox}
\subsection{Design Goals}
\label{sec:motivation-implications}

The three observations above suggest that single-GPU omni-model serving needs a scheduler that jointly controls request scheduling in time and GPU sharing in space. We summarize the design goals as follows.

\textbf{G1: Slack-based deadline protection.}
The scheduler should protect requests with little remaining latency slack, since the same shared-stage delay can be tolerable for one class but fatal for another.

\textbf{G2: Path rotation in the shared stage.}
The scheduler should regulate which execution paths enter the shared stage, instead of treating all requests as homogeneous token-progress units.

\textbf{G3: Bandwidth-guided SM throttling.}
The scheduler should use SM allocation as a bandwidth-control knob, selecting a bounded shared-stage cap that protects tight-SLO short-path requests without expanding the shared stage to the full GPU during co-running.

\section{HorizonServe}
\label{sec:method}

\subsection{System Overview}
\label{sec:design-overview}

\begin{figure}[!t]
    \centering
    \includegraphics[width=0.9\columnwidth]{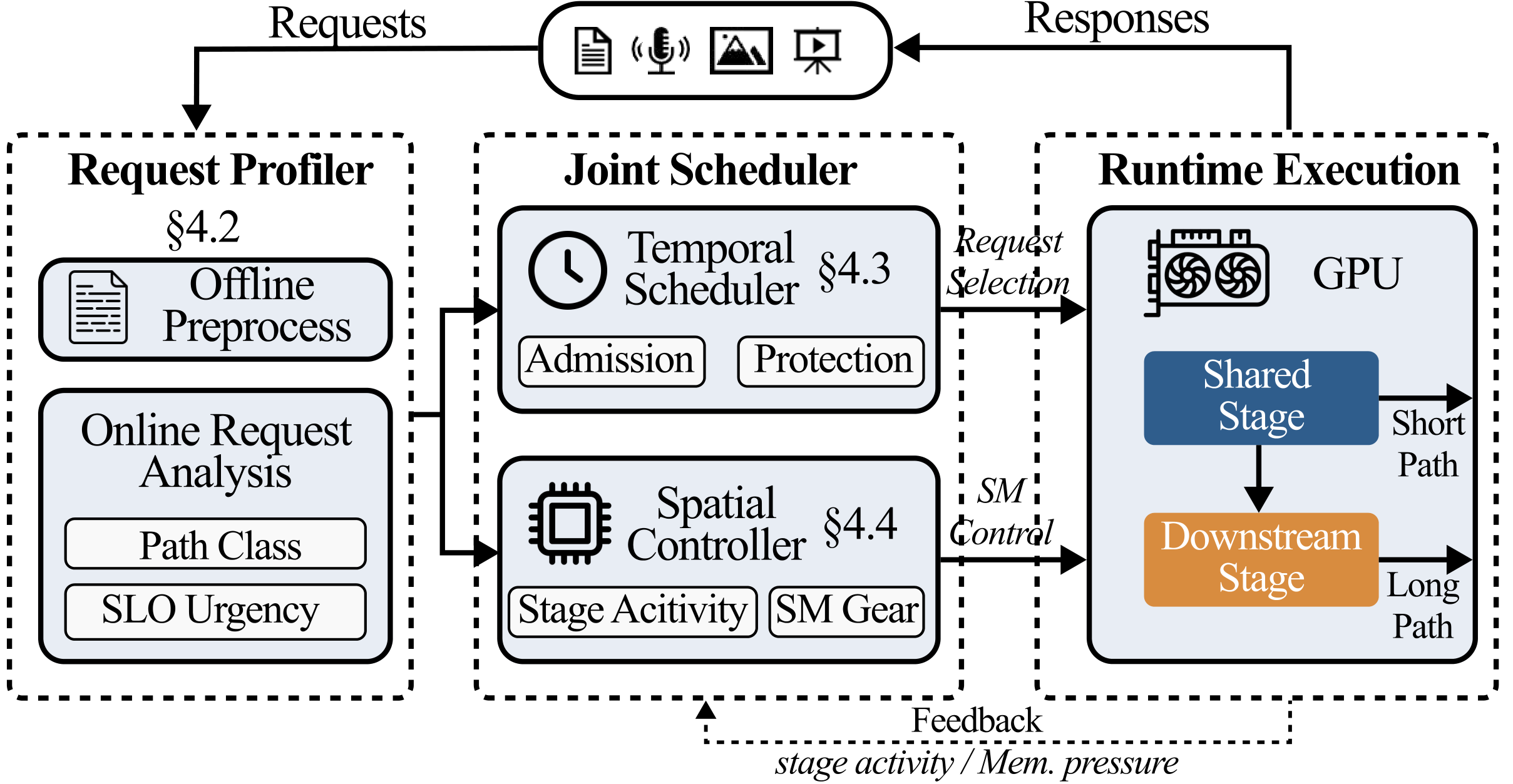}
    \caption{
    System overview of HorizonServe. 
    }
    \label{fig:HorizonServe-overview}
\end{figure}

To satisfy the design goals in Section~\ref{sec:motivation-implications}, we present HorizonServe, a multi-SLO inference serving system for single-GPU omni-model serving. Figure~\ref{fig:HorizonServe-overview} shows its workflow. HorizonServe separates the serving loop into four components: offline preprocessing, a request profiler, a temporal scheduler, and a spatial controller. These components share one control state rather than operating as independent optimizers.

Offline preprocessing builds a class-level SLO table and an SM ladder for each model-platform pair. At runtime, the request profiler maps each request to an execution path, deadline, remaining-slack signal, and downstream-activation flag. The temporal scheduler uses the path and slack signals to expose a temporary queue view to the base scheduler. The spatial controller uses downstream activity and memory-bandwidth pressure to select the shared-stage SM cap for the selected batch.

The three online components are coupled at every scheduling step. A short-path request can return after the shared stage, while a long-path request may activate downstream generation. Thus, admitting long-path work affects both current shared-stage queueing and later spatial pressure. HorizonServe therefore rotates shared-stage opportunities across paths, protects near-deadline requests, caps long-path bursts, and maps observed memory-bandwidth pressure to a bounded shared-stage SM cap during co-running.

The joint scheduler applies the queue-view and SM-cap decisions on the single GPU. The queue view determines which paths may consume shared-stage slots in this step, and the SM cap determines how the resulting batch shares the GPU with any active downstream stage. After execution, the runtime feeds back request progress, downstream activity, and memory-system usage. These feedback signals update both deadline slack and spatial pressure, closing the loop for the next batch and cap decision.

\subsection{Offline Preprocessing and Request Signals}
\label{sec:design-profiler}

The request profiler bridges offline preprocessing and online scheduling. Following common practice in SLO-driven LLM serving systems~\cite{298679,10.1145/3767295.3769315,10.1145/3774934.3786413}, HorizonServe constructs class-specific SLO targets and a bounded SM ladder before deployment. At runtime, the profiler maps each request to a class, assigns a deadline, and computes the slack signal used by the schedulers.

\textbf{Offline preprocessing.}
For each request class \(c \in \mathcal{C}\), HorizonServe measures solo p95 first-response latency \(p_{95}^{\mathrm{solo}}(c)\) and sets the class SLO target to \(\Delta_c=p_{95}^{\mathrm{solo}}(c)m_c\), where \(m_c\) is a deployment policy multiplier. It also sweeps fixed shared-stage SM caps, chooses the co-running cap \(S_{\mathrm{cap}}\) as the smallest SM count that reaches near-best decode TPOT without the full-allocation latency degradation observed in Section~\ref{sec:motivation-spatial}, and builds an SM ladder within \([S_{\min},S_{\mathrm{cap}}]\). Online serving only looks up these calibrated values; Section~\ref{sec:exp-setup} describes the SLO multiplier used in our experiments.

\textbf{Online request analysis.}
When request \(i\) arrives, HorizonServe maps it to a request class \(c(i)\) according to its input and requested output modalities. The class determines whether the request terminates after the shared stage or continues downstream, and gives the first-response deadline \(d_i=a_i+\Delta_{c(i)}\), where \(a_i\) is the arrival time.

\textbf{Remaining-slack ratio.}
At scheduling time, HorizonServe computes \(u_i(t)=(d_i-t)/\hat{r}_i(t)\), where \(d_i-t\) is the remaining time before the first-response deadline and \(\hat{r}_i(t)\) is a coarse remaining-time estimate rather than an oracle. The estimate combines offline per-stage profiles with online progress counters. The profile keys include input-token count, audio length, image or video count, and requested output path; the online counters subtract completed shared-stage decode progress and downstream remaining steps such as codec, waveform, or diffusion iterations. A request with \(u_i(t)>1\) has enough slack under the current estimate, while a request with \(u_i(t)\) close to 1 is near its deadline. HorizonServe marks a request as protected when \(u_i(t)<u_{\mathrm{guard}}\), using the guard band to absorb estimator error, scheduling granularity, and runtime variation.

The request profiler exposes the request path, the protected bit, and whether the request may activate downstream stages. These signals drive path rotation, bypass decisions, and later SM capping.

\subsection{Shared-Stage Path Rotation}
\label{sec:design-temporal}
\begin{figure}[!t]
    \centering
    \includegraphics[width=\columnwidth]{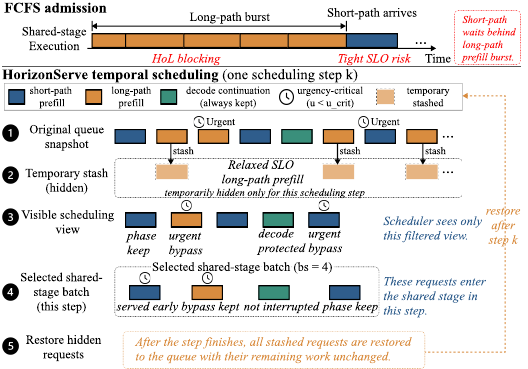}
\caption{
Shared-stage path rotation. HorizonServe hides ordinary non-phase prefill requests for one scheduling step, keeps protected and decode work visible, selects a shared-stage batch, and restores hidden work afterward.
}
    \label{fig:temporal-scheduling}
\end{figure}
The temporal scheduler consumes the path and slack signals produced by the request profiler and decides which requests participate in each shared-stage forward pass. This decision also shapes later spatial pressure: admitting a long-path request consumes a shared-stage slot now and may activate downstream work later. A temporal policy alone can protect queue order, but it cannot prevent admitted long-path work from creating bandwidth pressure for later batches. Therefore, HorizonServe rotates shared-stage opportunities while exposing downstream-activation signals to the spatial controller.

\textbf{Path rotation.}
HorizonServe uses an \(N{:}M\) rotation over shared-stage scheduling steps: step \(k\) is a short-path phase when \(k\bmod(N+M)<N\), and a long-path phase otherwise. During a short-path phase, ordinary long-path prefill requests are temporarily hidden from the base scheduler; during a long-path phase, ordinary short-path prefill requests can be hidden instead. Hidden requests are restored immediately after the step, so HorizonServe changes only the scheduler's view rather than permanently reordering the queue.

Figure~\ref{fig:temporal-scheduling} illustrates one short-path phase. Under FCFS admission, a long-path burst can occupy consecutive shared-stage slots. HorizonServe instead stashes ordinary non-phase prefills, so the base scheduler operates on a filtered view that favors the current phase.

\textbf{Slack guard.}
Path rotation alone can delay a request that is close to its deadline. HorizonServe therefore exempts protected requests from phase filtering: any request with \(u_i(t)<u_{\mathrm{guard}}\) remains visible even if its path does not match the current phase. Decode continuations also remain visible across phases, because hiding them can introduce token-streaming jitter and increase inter-token delay.

\textbf{Adaptive rotation ratio.}
The rotation ratio reflects current SLO pressure. HorizonServe fixes \(M=2\) rather than tuning it per workload. Here \(M\) is a shared-stage admission quota, not a downstream execution quota: downstream progress is governed by the downstream stage and the SM headroom selected by the spatial controller. The lower bound comes from feeding the shared-to-downstream pipeline. A long-path first response cannot begin output-side generation until its shared-stage work has been admitted; with only one long-path phase, that opportunity can be consumed by protected bypasses or decode continuations that remain visible across phases. Two long-path phases provide a small stable window for ordinary long-path prefills to enter the shared stage. We do not enlarge \(M\) further because long-path first-response SLOs are much looser, often by one to two orders of magnitude, than short-path SLOs (Figure~\ref{fig:m1-slo-spread}). The remaining elasticity is handled by the short-path count \(N\): HorizonServe measures the fraction of waiting short-path requests that are protected, maps it linearly to \(N\in[N_{\min},N_{\max}]=[2,8]\), and rounds it to the current short-path phase count. Even at maximum short-path pressure, long-path requests retain \(M/(N_{\max}+M)=20\%\) of scheduling steps; when pressure is low, \(N\) shrinks and their share increases.

\textbf{Long-path admission cap.}
HorizonServe further limits how many ordinary long-path prefills can occupy shared-stage slots. The cap prevents long-path bursts from consuming excessive KV-cache capacity, blocking short-path admission, and activating too many downstream generators at once. This is the temporal scheduler's direct handle on future spatial pressure: bounding long-path admission bounds how quickly downstream stages become active. When candidates exceed the cap, HorizonServe prioritizes earlier deadlines; protected requests remain exempt.

Together, path rotation, slack protection, and the long-path cap shape both current queueing delay and future downstream activity. The next section describes the complementary decision: how HorizonServe sets the shared-stage SM cap after those temporal choices create co-running pressure on the same GPU.

\subsection{Bandwidth-Guided SM Throttling}
\label{sec:design-spatial}

Given the batch stream produced by the temporal scheduler, the spatial controller controls how much GPU footprint the shared stage uses during execution. When long-path requests activate downstream stages, admission control alone is insufficient because co-running stages can increase memory-system pressure and delay short-path first responses. HorizonServe uses the shared-stage SM count as an indirect pressure-control knob: it does not claim that SMs are the bottleneck, but bounds the shared stage's execution footprint so full allocation does not amplify cross-stage contention. The cap is selected from the useful saturation range identified by fixed-SM profiling.

\begin{figure}[!t]
    \centering
    \includegraphics[width=0.90\columnwidth]{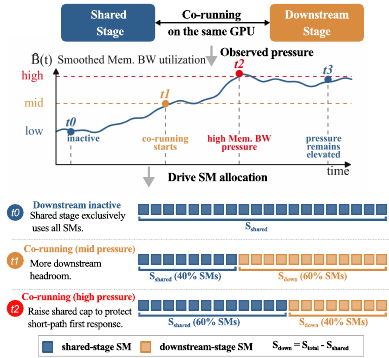}
\caption{
Bandwidth-guided SM throttling. HorizonServe uses an aggressive shared-stage cap when downstream stages are inactive and throttles the shared stage to a bounded ladder during co-running.
}
    \label{fig:spatial-control}
\end{figure}

\textbf{Bandwidth-pressure signal.}
HorizonServe tracks memory pressure using the normalized GPU memory-activity signal from PyNVML as an online proxy rather than a precise achieved-bandwidth counter, and smooths it with an EWMA \(\hat{B}(t)\) with weight \(\gamma\). Figure~\ref{fig:spatial-control} illustrates how this pressure signal drives a shared-stage cap and complementary downstream headroom.

\textbf{SM ladder.}
HorizonServe maps bandwidth pressure to a shared-stage SM cap through an offline ladder \(\mathcal{S}\), ordered from \(S_{\min}\) to \(S_{\mathrm{cap}}\). It normalizes \(\hat{B}(t)\) by the saturation pressure \(B_{\mathrm{sat}}\), clamps the ratio to \([0,1]\), and selects
\begin{equation}
s_{\mathrm{shared}}^{*}(t)=
\begin{cases}
S_{\mathrm{aggr}}, & A(t)=0,\\
\mathrm{snap}_{\mathcal{S}}\!\left(S_{\min}+\rho(t)(S_{\mathrm{cap}}-S_{\min})\right), & A(t)>0.
\end{cases}
\label{eq:spatial-cap}
\end{equation}
Here \(A(t)\) denotes downstream activity, \(S_{\mathrm{aggr}}\) is the aggressive no-downstream cap, and \(S_{\mathrm{cap}}<S_{\mathrm{aggr}}\le S_{\mathrm{tot}}\) is the maximum cap allowed during co-running. Thus, increasing pressure can only move the shared stage within the bounded ladder up to \(S_{\mathrm{cap}}\), never back to full allocation. This lets HorizonServe drain tight-SLO short-path requests while preserving downstream headroom and avoiding the harmful full-allocation regime from Section~\ref{sec:motivation-spatial}.

\textbf{Complementary downstream headroom.}
Given the total SM count \(S_{\mathrm{tot}}\), HorizonServe assigns the remaining budget \(s_{\mathrm{down}}(t)=S_{\mathrm{tot}}-s_{\mathrm{shared}}(t)\) as downstream headroom. This loop does not optimize the shared stage in isolation: it gives downstream generation headroom when pressure is low, but raises the shared stage toward \(S_{\mathrm{cap}}\) when bandwidth pressure threatens short-path first-response latency. The resulting progress or slowdown is reflected in the remaining-slack ratio used by the next temporal step.

When downstream stages are inactive, HorizonServe uses an aggressive shared-stage allocation. During co-running, low memory pressure leaves more SMs to downstream generation, while high pressure raises the shared-stage cap only up to \(S_{\mathrm{cap}}\), avoiding the full-allocation regime shown to be harmful in Section~\ref{sec:motivation-spatial}. To avoid oscillation, HorizonServe changes the cap only when the new ladder level differs from the previous cap by at least one ladder step \(\Delta_{\mathrm{step}}\).

This SM throttling policy complements path rotation. Rotation determines which requests enter the shared stage and thereby shapes downstream activity; the cap determines how much GPU parallelism the shared stage receives during the resulting co-running periods. The effect feeds back into later temporal decisions because memory pressure and execution slowdown change the remaining slack observed in subsequent steps.

\subsection{HorizonServe Scheduling Loop}
\label{sec:design-joint}

Sections~\ref{sec:design-temporal} and~\ref{sec:design-spatial} introduce temporal and spatial controls separately for clarity, but the runtime decision is joint. One scheduling step is a shared-stage opportunity: HorizonServe refreshes request and GPU signals, forms a temporary queue view, selects a batch, chooses the SM cap, and executes that batch. The queue view decides which long-path requests may create downstream work, while the SM cap determines how the admitted work shares the GPU with active downstream stages.

The dependence runs in both directions. Without spatial control, temporal admission can reserve shared-stage opportunities for the right paths, but the resulting batches may still be slowed by downstream bandwidth contention. Without temporal control, spatial throttling can react to memory pressure, but it cannot prevent long-path bursts from occupying shared-stage slots or decide which near-deadline requests should be protected before pressure appears. HorizonServe closes this loop by letting temporal admission shape future downstream activity and letting spatial control feed its progress and pressure back into the slack seen by the next step.

\begin{algorithm}[!t]
\caption{One HorizonServe Scheduling Step}
\label{alg:joint-scheduler}
\footnotesize
\begin{algorithmic}
\REQUIRE queue \(\mathcal{Q}\), step \(k\), SLO/control parameters, downstream activity \(A(t)\), memory-pressure signal \(B(t)\), previous cap \(s_{\mathrm{prev}}\)
\ENSURE batch \(\mathcal{B}\), shared-stage cap \(s_{\mathrm{shared}}\), downstream headroom \(s_{\mathrm{down}}\)
\STATE \textbf{Step 1: Refresh signals.} Update each request's class, path, deadline, slack, and protected bit from the SLO table and progress; sample the GPU memory-activity signal and update \(\hat{B}(t)\).
\STATE \textbf{Step 2: Build temporal view.} Update \(N\) from protected short-path pressure, choose phase \(\phi\), hide ordinary non-\(\phi\) prefills, keep protected and decode work visible, cap ordinary long-path prefills, and call \(\mathcal{B}\leftarrow\mathrm{BaseScheduler}(\mathcal{V})\).
\STATE \textbf{Step 3: Select SM split.} Compute \(s^{*}\) from Equation~\ref{eq:spatial-cap}. Apply it if downstream stages are inactive or the ladder level changes by \(\Delta_{\mathrm{step}}\); otherwise keep \(s_{\mathrm{prev}}\). Set \(s_{\mathrm{down}}\leftarrow\max(0,S_{\mathrm{tot}}-s_{\mathrm{shared}})\).
\STATE \textbf{Step 4: Execute and publish.} Run \(\mathcal{B}\) with cap \(s_{\mathrm{shared}}\), restore hidden requests, and publish request progress, downstream activity, and \(\hat{B}(t)\).
\RETURN \(\mathcal{B}, s_{\mathrm{shared}}, s_{\mathrm{down}}\)
\end{algorithmic}
\end{algorithm}

Algorithm~\ref{alg:joint-scheduler} shows this loop. Step~1 refreshes the request and GPU signals left by prior decisions. Step~2 builds the temporal view, including the long-path cap that bounds future downstream work. Step~3 chooses the SM split for the selected batch, trading shared-stage progress against downstream headroom during co-running. Step~4 publishes progress and pressure so the next step observes the consequences of both decisions. HorizonServe therefore controls admission and SM allocation as one feedback loop, while the base scheduler still performs low-level batch construction and resource checks.

\section{Implementation}
\label{sec:method-implementation}

We implement HorizonServe on top of vLLM-Omni~\cite{yin2026vllmomnifullydisaggregatedserving}. HorizonServe keeps vLLM-Omni's batching, KV-cache management, and model execution unchanged, and interposes only at scheduling-step boundaries. For each step, it annotates requests with class and deadline metadata: the request class is inferred from a priority field set at the API layer ($0$~=~text output, $10$~=~audio output, $20$~=~image output), and the deadline is stamped at admission time using a monotonic clock with a per-type SLO threshold read from environment variables, so thresholds can be recalibrated to measured solo-run latencies on the target device without recompilation. HorizonServe then hides requests that should not be visible in the current temporary scheduler view, invokes the original vLLM-Omni scheduler, and restores the hidden requests. This realizes HorizonServe's admission policy without moving KV-cache state or rewriting batch construction.

For spatial control, our prototype applies the selected shared-stage SM cap with CUDA Green Contexts. Each stage process pre-allocates a full ladder of Green Contexts at startup, one per SM level, so runtime control only selects an existing level and activates the corresponding context before a forward pass. Stage~0 and Stage~1 hold complementary ladders: their SM counts always sum to the physical GPU total, ensuring the hardware is fully utilized under any SM split. HorizonServe can also use any backend that exposes a per-stage SM cap. Stage processes pass compact control state through a shared-memory monitor in \texttt{/dev/shm}; the record carries downstream-stage activity and the current SM decision. Stage outputs are exchanged through the runtime's existing inter-process queues. We sample the GPU memory-activity signal with PyNVML, the Python bindings for the NVIDIA Management Library (NVML), at scheduling-step boundaries and publish the normalized value through the same monitor for the spatial controller. 

\section{Experiments}
\label{sec:experiments}

\subsection{Experimental Setup}
\label{sec:exp-setup}

\textbf{Platform.}
All experiments use single-GPU co-location: the shared stage and downstream stages run as separate processes on the same GPU. Table~\ref{tab:exp-platform} summarizes the host and GPU environments. Our evaluation spans two single-GPU platforms, RTX 6000 Ada and RTX PRO 6000, allowing us to study the system under different SM counts, memory capacities, and memory bandwidths. Both platforms use Ubuntu 22.04.5 LTS, CUDA 13.1, and NVIDIA driver 590.48.01.

\begin{table}[!t]
\centering
\caption{Experimental GPU platforms.}
\label{tab:exp-platform}
\scriptsize
\setlength{\tabcolsep}{2.2pt}
\begin{tabular*}{\columnwidth}{@{}p{0.24\columnwidth}@{\extracolsep{\fill}}p{0.34\columnwidth}p{0.34\columnwidth}@{}}
\toprule
\textbf{Item} & \textbf{RTX 6000 Ada} & \textbf{RTX PRO 6000} \\
\midrule
Host CPU & AMD Ryzen Threadripper PRO 5955WX & 24 vCPU Intel Xeon Platinum 8568Y+ \\
System memory & 128\,GB  & 120\,GB \\
GPU memory & 48\,GB GDDR6 ECC & 96\,GB GDDR7 ECC \\
Peak bandwidth & 960\,GB/s & 1,792\,GB/s \\
Compute resources & 142 SMs & 188 SMs \\
\bottomrule
\end{tabular*}
\end{table}
\begin{figure*}[!t]
    \centering
    \includegraphics[width=0.92\textwidth]{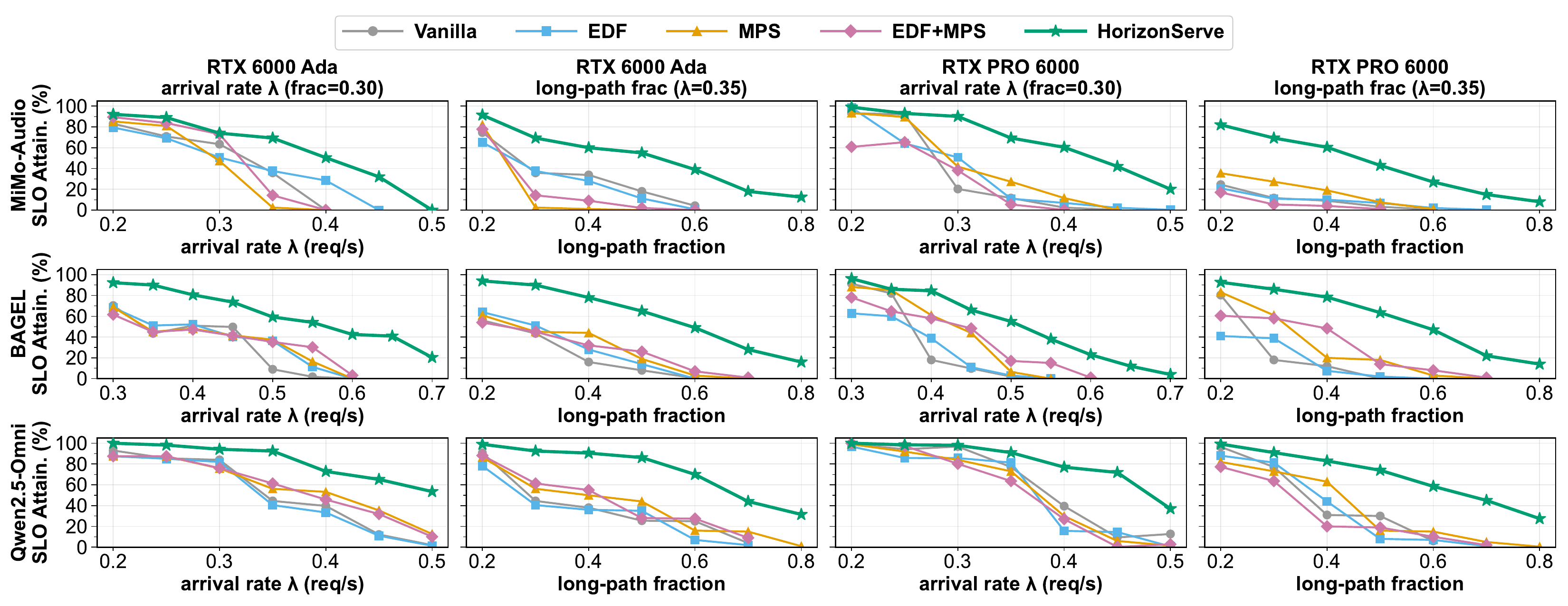}
    \caption{
    End-to-end SLO attainment under increasing arrival rates and long-path request ratios.
    }
    \label{fig:e2e-master}
\end{figure*}
\textbf{Models.}
We evaluate three omni-model serving pipelines: Qwen2.5-Omni~\cite{xu2025qwen25omnitechnicalreport}, MiMo-Audio~\cite{coreteam2025mimoaudioaudiolanguagemodels}, and BAGEL~\cite{deng2025bagel}. They cover distinct downstream-generation patterns: speech-output generation, audio-code generation, and image generation. Together, they exercise both short paths that terminate after the shared stage and long paths that activate downstream stages before producing the first user-visible output.

\textbf{Workloads.}
Requests arrive according to a Poisson process. We construct model-specific request classes from the modalities supported by each model, using ShareGPT conversations for interactive text prompts, LibriSpeech for audio inputs~\cite{panayotov2015librispeech}, Food-101 for image inputs~\cite{bossard2014food}, UCF-101-style synthetic clips for video inputs~\cite{soomro2012ucf101}, and CNN/DailyMail for long-text summarization prompts~\cite{hermann2015teaching}. Each experiment uses a mixed stream of short-path and long-path requests and runs each plotted point for 15 minutes. We observed consistent trends across repeated runs and report representative runs. In arrival-rate sweeps, we increase the total offered load at a fixed traffic mix. In long-path sweeps, we keep the offered load fixed and vary the fraction of requests that activate downstream stages. This separates queueing pressure from downstream co-running pressure.

\textbf{Baselines.}
We use vLLM-Omni (v0.16.0) as the base runtime and vanilla baseline because it supports the omni-model pipelines evaluated in this paper and exposes their shared/downstream execution structure on a single GPU. Existing VLM/LMM systems mainly target input-side multimodal processing followed by text generation. GPU-sharing systems provide low-level mechanisms rather than omni-path admission policies, and disaggregated serving systems assume multi-GPU or cluster-level stage placement rather than single-GPU co-location~\cite{guo2025rserve,10.1145/3772052.3772254,bai2026epd,yin2026vllmomnifullydisaggregatedserving,nanoflow,bullet,muxserve2024}. We therefore build controlled baselines inside the same vLLM-Omni runtime: \emph{Vanilla}, \emph{EDF}, \emph{MPS}, and \emph{EDF-MPS}. Keeping the runtime fixed makes the comparison focus on scheduling and GPU-sharing policy rather than model support, batching implementation, or IPC differences. \emph{Vanilla} is unmodified vLLM-Omni. \emph{EDF} orders eligible requests by first-response deadline while leaving GPU sharing unchanged. \emph{MPS} is our practical spatial-sharing baseline: it delegates multi-process GPU sharing to NVIDIA's runtime-managed MPS scheduler, allowing shared and downstream stage processes to co-run on one GPU without application-level SM control. \emph{EDF-MPS} combines EDF request ordering with MPS-based runtime GPU sharing. Together, these baselines isolate deadline ordering, runtime-managed GPU co-running, and their direct combination without HorizonServe's feedback between path-aware admission and stage-level GPU allocation.

\textbf{Metrics.}
HorizonServe's objective is to increase SLO-satisfied goodput, i.e., useful throughput from requests that meet their class-specific SLOs. In the main result, we report SLO attainment: the fraction of requests whose user-visible response satisfies the target. At a fixed offered load, higher SLO attainment directly means higher SLO-satisfied goodput. Because omni requests expose different output forms, we report the first response latency, which includes the queueing time. The latency metric used to judge attainment is class-specific: text-output requests use time-to-first-token (TTFT), audio-output requests use time-to-first-output (TTFO), and image-output requests use end-to-end latency. We set each class SLO target to \(5\times\) its isolated p95 first-response latency, following prior SLO and goodput-oriented serving practice~\cite{298679,10.1145/3767295.3769315,10.1145/3774934.3786413}. We use the same multiplier for all request classes, models, GPUs, and serving policies. This keeps SLO tightness tied to each class's solo latency while using a common policy across workloads. We compute SLO attainment over completed requests in each run and also report p95 response latency, shared-stage TPOT, GPU memory-bandwidth utilization, and scheduling overhead.

\subsection{Main Results}
\label{sec:exp-e2e}

Figure~\ref{fig:e2e-master} shows the main result under mixed omni traffic. For each model and GPU platform, we run two sweeps: increasing the total Poisson arrival rate at a fixed traffic mix, and increasing the long-path request ratio at a fixed offered load. SLO attainment is computed with the class-specific first-response metric from Section~\ref{sec:exp-setup}.

HorizonServe consistently preserves higher SLO attainment in the stressed region of both sweeps. The largest gap appears on RTX PRO 6000 with Qwen2.5-Omni: at arrival rate 0.45, HorizonServe reaches 71.9\% SLO attainment, while the best baseline reaches 14.6\%. The same trend appears on RTX 6000 Ada, where HorizonServe keeps Qwen2.5-Omni at 53.5\% attainment at arrival rate 0.5, compared with 12.7\% for the best baseline. Thus, the larger PRO platform raises absolute capacity but does not remove the scheduling problem when the mixed stream reaches the stressed region; uncontrolled sharing still loses most SLO-satisfied work.

The long-path-ratio sweep stresses the other half of the design. It keeps the total offered load fixed while increasing the amount of downstream generation. Under 60\% long-path traffic on RTX 6000 Ada, HorizonServe reaches 70.0\% attainment for Qwen2.5-Omni and 49.0\% for BAGEL, compared with 27.5\% and 7.0\% for the best baselines. This shows that the gain is not only from handling higher arrival rates; it also comes from preventing downstream-heavy traffic from consuming shared-stage opportunities and creating cross-stage contention.

The baselines fail in complementary ways. EDF improves deadline ordering, but it does not reduce the downstream work created by long-path bursts or change how co-running stages share the GPU. MPS improves concurrent execution, but it keeps the original request admission policy and therefore cannot protect short-path slack. EDF-MPS combines both mechanisms, yet still lacks a feedback loop between path admission and stage-level GPU allocation. HorizonServe performs better because the temporal decision and the spatial decision are made together: it controls which paths enter the shared stage and how much SM capacity the shared stage uses when downstream stages are active.

\subsection{Latency and Per-Class Analysis}
\label{sec:exp-breakdown}

\begin{figure}[!t]
    \centering
    \includegraphics[width=0.95\columnwidth]{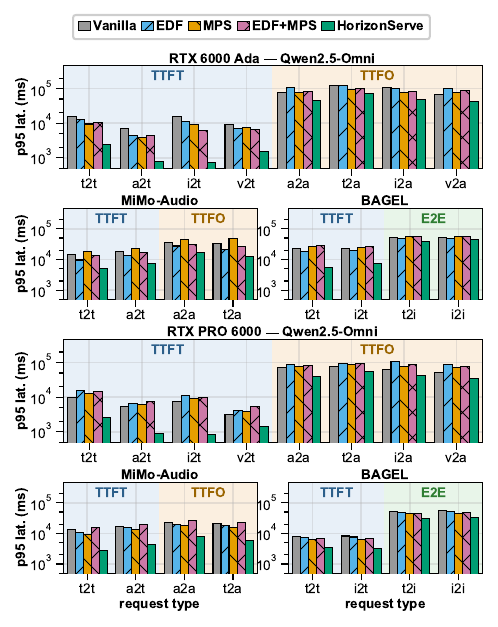}
    \caption{
    Per-class p95 response latency including queueing delay.
    }
    \label{fig:perclass-latency}
\end{figure}

This section examines whether the end-to-end gains in Figure~\ref{fig:e2e-master} come from broad request-level improvement rather than one dominant class. Figure~\ref{fig:perclass-latency} breaks down p95 response latency by request class at representative load points. The latency is measured from request arrival to the first user-visible response, so it includes queueing delay as well as model execution. Since different classes expose different response signals, we report TTFT for text-output requests, TTFO for audio-output requests, and end-to-end latency for image-output requests.

The per-class view complements the aggregate SLO attainment in Section~\ref{sec:exp-e2e}. A serving policy can improve the aggregate result by favoring an easier request class, but HorizonServe aims to protect tight-SLO short-path requests while still admitting long-path work. Compared with the best baseline for each class, HorizonServe reduces p95 response latency by 57.1\% for Qwen2.5-Omni, 42.1\% for MiMo-Audio, and 38.4\% for BAGEL on RTX 6000 Ada. The gains persist on RTX PRO 6000, where the corresponding p95 reductions are 55.2\%, 63.7\%, and 38.4\%. These reductions cover both short-path and long-path classes, indicating that HorizonServe is not simply shifting delay from one request type to another.

The latency breakdown also explains why EDF and MPS are insufficient alone. EDF can reduce queueing for some near-deadline classes, but it leaves co-running interference unresolved. MPS can reduce some long-path latency by allowing concurrent kernels, but it does not prevent long-path work from occupying shared-stage opportunities. HorizonServe reduces both effects: path rotation and slack protection lower shared-stage waiting time for tight-SLO classes, while bandwidth-guided SM throttling reduces interference when downstream stages are active.

We also report shared-stage (Stage~0) TPOT in Table~\ref{tab:stage0-tpot}. The table reports p50/p95 TPOT, checking whether improving first-response latency harms token streaming in the shared stage, which is the component directly affected by HorizonServe's admission and SM-allocation decisions.

\begin{table}[!t]
\centering
\caption{Shared-stage (Stage~0) TPOT by serving policy. Values are p50/p95 in ms/token.}
\label{tab:stage0-tpot}
\scriptsize
\begin{tabular*}{\columnwidth}{@{}l@{\extracolsep{\fill}}lccccc@{}}
\toprule
\textbf{Model} & \textbf{GPU} & \textbf{Van.} & \textbf{EDF} & \textbf{MPS} & \textbf{E+M} & \textbf{Horizon} \\
\midrule
Qwen2.5-Omni & Ada & 39.5/62.9 & 39.7/63.3 & 39.7/45.9 & 39.8/45.8 & 40.0/64.0 \\
 & PRO & 36.0/57.5 & 36.1/57.8 & 36.1/42.0 & 36.2/41.8 & 36.3/58.1 \\
\midrule
MiMo-Audio & Ada & 52.4/54.6 & 52.4/54.7 & 52.5/55.0 & 52.4/54.8 & 52.4/54.6 \\
 & PRO & 48.0/50.0 & 48.0/50.1 & 48.1/50.4 & 48.0/50.2 & 48.0/50.0 \\
\midrule
BAGEL & Ada & 37.1/38.6 & 37.1/38.6 & 37.1/38.7 & 37.1/38.7 & 37.4/38.8 \\
 & PRO & 34.1/35.5 & 34.0/35.5 & 34.1/35.6 & 34.1/35.6 & 34.0/35.4 \\
\bottomrule
\end{tabular*}
\end{table}

Table~\ref{tab:stage0-tpot} shows that HorizonServe's first-response gains do not come from sacrificing shared-stage token streaming. Relative to Vanilla, HorizonServe changes p50 TPOT by at most 0.5\,ms/token on RTX 6000 Ada and 0.3\,ms/token on RTX PRO 6000. The p95 TPOT is also essentially unchanged for MiMo-Audio and BAGEL, and changes by only about 1\,ms/token for Qwen2.5-Omni. MPS sometimes lowers Qwen2.5-Omni p95 TPOT, but Figure~\ref{fig:e2e-master} shows that better token streaming alone does not preserve SLO-satisfied goodput without path and slack control.

\subsection{Hardware-Signal Analysis}
\label{sec:exp-hardware}

\begin{figure}[!t]
    \centering
    \includegraphics[width=0.95\columnwidth]{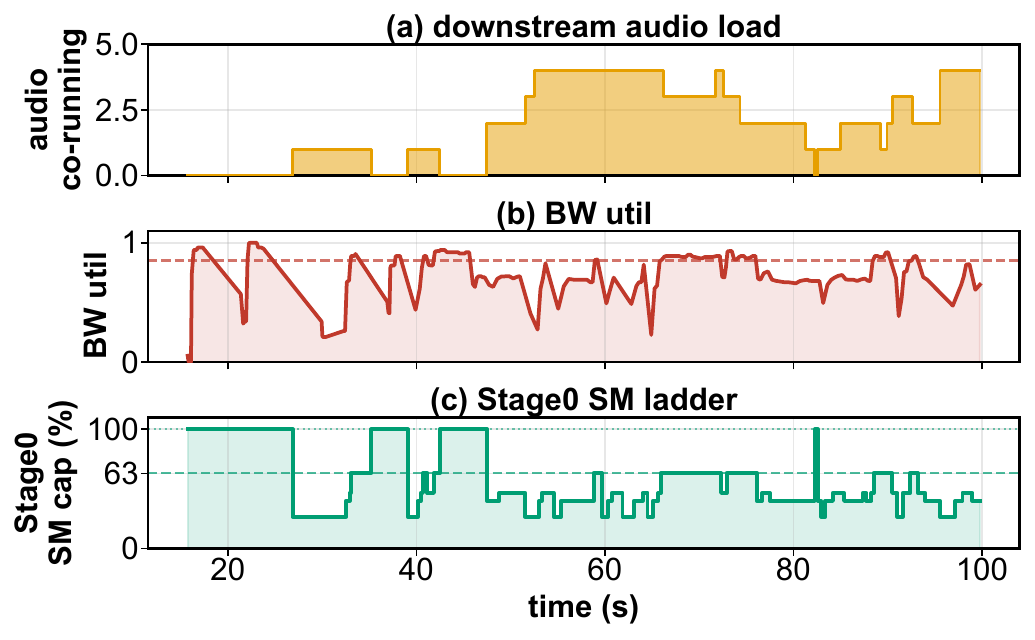}
    \caption{
    Hardware signal and SM response in a representative Qwen2.5-Omni run.
    }
    \label{fig:hardware-signal}
\end{figure}

Figure~\ref{fig:hardware-signal} shows the spatial-control loop for Qwen2.5-Omni on RTX 6000 Ada at \(\lambda=0.3\) with 40\% audio traffic. The top panel shows the number of concurrent audio-output requests, which marks when downstream generation is active. The middle panel shows the memory-pressure proxy sampled through PyNVML and smoothed with an exponentially weighted moving average (\(\gamma=0.4\)). The bottom panel shows the shared-stage SM cap selected by HorizonServe.

The three traces align in time. When no audio request is running, the shared stage can use its aggressive cap because there is no downstream co-running stage. Once audio generation becomes active, HorizonServe switches to the calibrated ladder. Under low or moderate pressure, it leaves more SM headroom to downstream generation. As measured memory pressure grows and often crosses the saturation threshold, the controller raises the shared-stage cap toward \(S_{\mathrm{cap}}\) so short-path requests can still reach their first response under tighter SLOs. Across this run, the correlation between the smoothed memory signal and the selected cap is \(0.926\), showing that the controller tracks the pressure signal while staying within the cap range motivated by Figure~\ref{fig:motivation-spatial}.

This figure is not a standalone performance comparison or a direct achieved-bandwidth measurement; it validates the online pressure proxy used by the controller. The PyNVML signal follows downstream activity and drives a stable bounded-cap response after EWMA smoothing, linking the fixed-SM profiling in Section~\ref{sec:motivation-spatial} to the runtime policy in Section~\ref{sec:design-spatial}.

\subsection{Ablation and Overhead}
\label{sec:exp-ablation}

\begin{figure}[!t]
    \centering
    \includegraphics[width=\columnwidth]{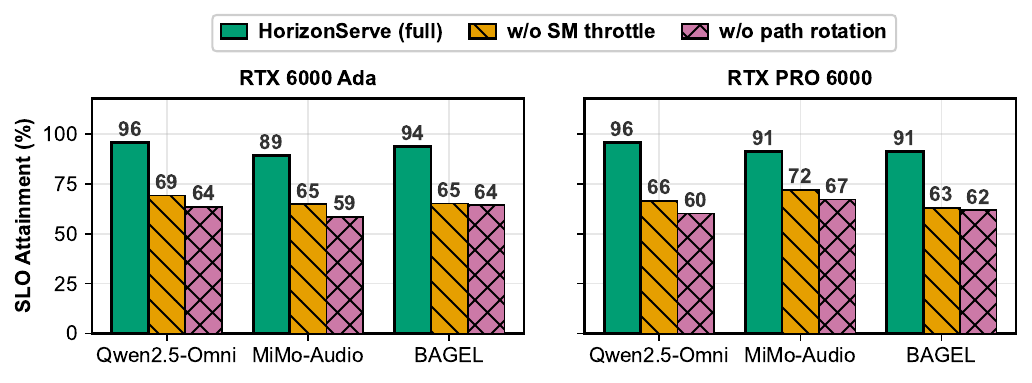}
    \caption{
    Ablation of HorizonServe mechanisms under representative mixed workloads.
    }
    \label{fig:ablation}
\end{figure}

This section isolates the contribution of the mechanisms shown in our ablation and quantifies runtime overhead. Figure~\ref{fig:ablation} compares full HorizonServe with variants that remove shared-stage path rotation or bandwidth-guided SM throttling while keeping the remaining policy unchanged. We run these variants on both GPU platforms at representative load points from the main experiment. The figure reports SLO attainment, so lower bars indicate that the removed mechanism is necessary for preserving SLO-satisfied goodput under mixed traffic.

The ablation confirms that both sides of the joint policy are needed. Removing path rotation lets long-path bursts occupy shared-stage opportunities, which primarily hurts tight-SLO short paths and also increases later downstream pressure. Removing bandwidth-guided SM throttling leaves the temporal scheduler intact, but exposes requests to co-running interference once downstream generation is active. Full HorizonServe performs best because the two mechanisms close the loop: temporal admission controls future downstream work, and SM throttling controls how active downstream work shares the GPU with the shared stage.

\begin{table}[!t]
\centering
\caption{Runtime overheads introduced by HorizonServe.}
\label{tab:overhead}
\scriptsize
\begin{tabular}{@{}p{0.31\columnwidth}p{0.25\columnwidth}p{0.11\columnwidth}p{0.11\columnwidth}@{}}
\toprule
\textbf{Component} & \textbf{Metric} & \textbf{Ada} & \textbf{PRO} \\
\midrule
Request annotation & \(\mu\)s/request & 0.4 & 0.4 \\
Scheduler-view wrapper & \(\mu\)s/step & 1.0 & 1.0 \\
PyNVML sampling & \(\mu\)s/step & 1.3 & 1.4 \\
SM-cap selection & \(\mu\)s/step & 0.3 & 0.3 \\
Shared-memory monitor & \(\mu\)s/update & 33 & 35 \\
Green Context activation & \(\mu\)s/switch & 4.8 & 5.0 \\
End-to-end tax & Throughput loss (\%) & \(<1\) & \(<1\) \\
\bottomrule
\end{tabular}
\end{table}

Table~\ref{tab:overhead} reports the online costs on the serving path. Request annotation, scheduler-view filtering, PyNVML sampling, and SM-cap selection are all within a few microseconds. The shared-memory monitor is the largest Python-side control cost, but it is still tens of microseconds per update, and Green Context activation adds about \(5\,\mu\)s per switch. The measured end-to-end throughput loss is below 1\% on both platforms, so the runtime control loop is lightweight relative to model execution.

Offline preprocessing runs before serving starts. SLO collection runs each request class in isolation, measures p95 solo first-response latency, and stores the resulting SLO table. In our setup, Qwen2.5-Omni has eight request classes and takes about 30--60 minutes, with short text classes taking about 10 seconds and longer audio or video-related classes taking 2--5 minutes each. MiMo-Audio has four classes and takes about 10 minutes. BAGEL has four classes and takes about 15 minutes, with the diffusion-output class taking about 3 minutes. HorizonServe also performs the fixed-SM sweep used to set the SM ladder and \(S_{\mathrm{cap}}\). These offline artifacts are regenerated only when the model, hardware platform, or SLO policy changes.

\section{Related Work}
\label{sec:related-work}

\textbf{Model serving systems.}
General prediction-serving systems optimize batching, model selection, autoscaling, interference control, and predictable latency for DNN inference~\cite{clipper2017,mark2019,clockwork2020,nexus2019,infaas2021,shepherd2023,usher2024,11223102}. SLO-aware and multi-SLO serving has been studied through latency-targeted scheduling, goodput optimization, request prioritization, speculative decoding, and fine-grained execution control~\cite{mark2019,clockwork2020,shepherd2023,298687,mooncake2024,llumnix2024,10.1145/3767295.3769315,10.1145/3774934.3786413}. Recent LLM serving systems improve iteration-level batching, KV-cache management, and structured generation~\cite{280922,vllm,sglang2024}; they also optimize chunked prefill, prefill--decode disaggregation, phase splitting, and KV-cache-centric serving~\cite{298679,298687,10.1109/ISCA59077.2024.00019,mooncake2024,pdserve2024}. Other work addresses long-context scheduling, request migration, and serverless loading~\cite{10.1145/3694715.3695948,llumnix2024,serverlessllm2024}. FlexGen and PowerInfer study limited-resource or consumer-GPU LLM inference~\cite{sheng2023flexgen,song2024powerinfer}. These systems are closest in their concern for latency and resource efficiency, but they primarily target text generation or model-level serving. HorizonServe instead targets multimodal, multi-SLO inference serving: it treats an omni request as a shared stage followed by optional output-side generators, and schedules these paths under first-response SLOs.

\textbf{Multimodal serving and GPU sharing.}
Multimodal and generative serving systems extend the serving target beyond text by optimizing visual encoders, multimodal preprocessing, encode-prefill-decode pipelines, or disaggregated any-to-any serving~\cite{guo2025rserve,10.1145/3772052.3772254,bai2026epd,yin2026vllmomnifullydisaggregatedserving,ye2026genserveefficientcoservingheterogeneous}. GPU sharing mechanisms and systems provide the substrate for co-location. They include CUDA streams, MPS, and MIG~\cite{nvidia-cuda-programming-guide,nvidia-mps,nvidia-mig}, fast GPU sharing and preemption~\cite{salus2020,reef2022}, spatio-temporal partitioning and hardware compute partitioning~\cite{gpulet2022,bakita2023hardware,nvidia-green-contexts}, and LLM-oriented spatial sharing~\cite{nanoflow,bullet,muxserve2024}. These works provide important mechanisms, but they do not jointly handle output-side omni paths, heterogeneous first-response SLOs, single-GPU shared/downstream co-location, and feedback between path admission and stage-level GPU allocation.

\section{Conclusion}
\label{sec:conclusion}

Omni-model serving introduces heterogeneous execution paths through a shared multi-stage pipeline. This path heterogeneity creates temporal interference in the shared backbone and spatial interference across concurrent GPU stages. We presented HorizonServe, a joint temporal-spatial scheduler that rotates shared-stage opportunities across paths and throttles the shared stage during co-running. Across three omni-model workloads and two GPU platforms, HorizonServe improves SLO attainment by up to 4.9$\times$ in arrival-rate sweeps and 7.0$\times$ under downstream-heavy traffic. It also reduces per-class first-response latency by 38.4--63.7\%, preserves shared-stage TPOT, and adds less than 1\% throughput overhead. These results show that deadline slack and bandwidth pressure must be controlled together for efficient multi-SLO omni-model serving.

\bibliographystyle{ACM-Reference-Format}
\bibliography{sample-base}

\end{document}